\documentstyle[pra,aps,twocolumn,epsf]{revtex}

\def\H{{\cal H}}\def\K{{\cal K}}
\def\D{{\cal D}}
\def\eps{\varepsilon}
\def\co{{\rm co}}
\def\pr{{\bf P}}
\def\dU{\!dU\ }
\def\ERE{E_{\rm RE}}
\def\EoF{E_{\rm F}}

\newcommand{\F}{{I\hskip-.39em F}}
\def\flip{\F}
\def\ft{\widehat f}
\def\flipt{\widehat\flip}

\def\til#1{{\widetilde#1}}

\def\Ubar{\overline{U}}

\def\Rplus{\Rl\cup\lbrace+\infty\rbrace}

\def\st{{\cal S}}
\def\examp#1 #2.{\par\vskip12pt\noindent{\sl Example #1: #2.}\newline\vskip2pt\noindent}
\def\UU{UU}
\def\UUb{U\overline U}
\def\OO{OO}
\def\sep{{\cal D}}
\def\ptr{\Theta_2}
\def\surep{{\cal D}}
\def\SU#1{SU_{#1}}

\def\idty{{\leavevmode{\rm 1\ifmmode\mkern -5.4mu\else\kern -.3em\fi I}}}
\def\Ibb #1{ {\rm I\ifmmode\mkern -3.6mu\else\kern -.2em\fi#1}}
\def\Ird{{\hbox{\kern2pt\vbox{\hrule height0pt depth.4pt width5.7pt
    \hbox{\kern-1pt\sevensy\char"36\kern2pt\char"36} \vskip-.2pt
    \hrule height.4pt depth0pt width6pt}}}}
\def\Irs{{\hbox{\kern2pt\vbox{\hrule height0pt depth.34pt width5pt
       \hbox{\kern-1pt\fivesy\char"36\kern1.6pt\char"36} \vskip -.1pt
       \hrule height .34 pt depth 0pt width 5.1 pt}}}}

\def\ibb #1{\leavevmode\hbox{\kern.3em\vrule
     height 1.5ex depth -.1ex width .2pt\kern-.3em\rm#1}}
 \def\Cx {{\ibb C}} \def\Rl {{\Ibb R}}
\def\SS{{\leavevmode\hbox{\kern.3em
        \vrule  height 1.5ex depth -.8ex width .6pt\kern .05em
        \vrule  height .7ex depth   0 ex width .6pt\kern-.35em
        \rm S}}}  
\def\SS{{\leavevmode\hbox{\kern.3em
        \vrule  height 1.6ex depth -.8ex width .5pt\kern .09em
        \vrule  height .7ex depth   0 ex width .5pt\kern-.33em
        \rm S}}}  

\def\bra #1{\langle #1\vert}
\def\ket #1{\vert #1\rangle}
\def\braket #1#2{\langle #1 \vert #2\rangle}
\def\ketbra #1#2{\vert #1\rangle \! \langle #2\vert}
\def\abs #1{\vert#1\vert}

\def\QED{\leavevmode\unskip\penalty9999 \hbox{}\nobreak\hfill
     \quad\hbox{\leavevmode  \hbox to.77778em{%
               \hfil\vrule   \vbox to.675em%
               {\hrule width.6em\vfil\hrule}\vrule\hfil}}
     \par\vskip24pt}
\def\norm #1{\Vert #1\Vert}
\def\tr{{\rm tr}}
\def\Set#1#2#3{#1\lbrace#2#1\vert#3#1\rbrace}

\newcommand\beq{\begin{equation}}
\newcommand\eeq{\end{equation}}
\newcommand\bea{\begin{eqnarray}}
\newcommand\eea{\end{eqnarray}}

\title{Entanglement Measures under Symmetry}
 \author{K.G. ~H. Vollbrecht\thanks{Electronic Mail:
\tt{k.vollbrecht@tu-bs.de}}{{}\quad and\ }
  R.~F. Werner\thanks{Electronic Mail: \tt{R.Werner@tu-bs.de}}
  \\[1ex]
  {\small Institut f{\"u}r Mathematische Physik, TU Braunschweig,}\\
  {\small Mendelssohnstr. 3, 38106 Braunschweig, Germany.}}
\date{\today}

\begin{document}
\maketitle

\begin{abstract} We show how to simplify the computation of the 
entanglement of formation and the relative entropy of entanglement 
for states, which are invariant under a group of local symmetries. 
For several examples of groups we characterize the state spaces, 
which are invariant under these groups. For specific examples we 
calculate the entanglement measures. In particular, we derive an 
explicit formula for the entanglement of formation for $U\otimes 
U$-invariant states, and we find a counterexample to the 
additivity conjecture for the relative entropy of entanglement. 
\end{abstract} 


\narrowtext

\section{Introduction}
One of the reasons the general theory of entanglement has proved
to be so difficult is the rapid growth of dimension of the state
spaces. For bipartite entanglement between $d_1$- and
$d_2$-dimensional Hilbert spaces, entanglement is a geometric
structure in the $(d_1^2d_2^2-1)$-dimensional state space. Hence
even in the simplest non-trivial case ($d_1=d_2=2$; 15 dimensions)
naive geometric intuitions can be misleading. On the other hand,
the rapid growth of dimensions is partly responsible for the
potential of quantum computing. Hence exploring this complexity is
an important challenge for quantum information theory.

Model studies have been an important tool for developing and
testing new concepts and relations in entanglement theory, both
qualitative and quantitative. In this paper we explore a method
for arriving at a large class of models, which are at the same
simple, and yet show some of the interesting features of the full
structure.

The basic idea, namely looking at sets of states which are
invariant under a group of local unitaries is not new, and goes
back to the first studies of entanglement \cite{Wer,Pop} in the
modern sense. Two classes, in particular, have been considered
frequently: the so-called {\it Werner states} (after\cite{Wer}),
which are invariant under all unitaries of the form $U\otimes U$,
and the so-called {\it isotropic states} \cite{isotropic}, which
are invariant under all $U\otimes \overline U$, where $\overline
U$ is the complex conjugate of $U$ in some basis. Symmetry has
also been used in this way  to study tripartite entanglement
\cite{tripartite},\cite{tripartiteC}. A recent paper of Rains
\cite{Rains2} discusses distillible entanglement under symmetry,
so we have eliminated the pertinent remarks from this paper.

Several of the ingredients of our general theory, for example the
role of the twirl projection and the commutant, have been noted in
these special cases and can be considered to be well-known. The
computation of the relative entropy of entanglement \cite{ERE} was
known \cite{EREW} for Werner states.  The first study in which
symmetry is exploited to compute the entanglement of formation
\cite{EoF} beyond the Wootters formula \cite{Wootters} is
\cite{TerVol}, where the case of isotropic states is investigated.
Our theory of entanglement of formation can be viewed as an
abstract version of arguments from that paper.

What is new in the present paper is firstly the generality. We
regard our theory as a toolkit for constructing examples adapted
to specific problems, and we have tried to present it in a
self-contained way facilitating such applications. Exploring all
the possibilities would have been too much for a single paper but,
of course, we also have some new results in specific examples.

Our most striking specific result is perhaps a counterexample to
the conjecture that the relative entropy of formation should be
additive. The evidence in favor of this conjecture had been partly
numerical, but it was perhaps clear that a random search for
counterexamples was not very strong evidence to begin with: the
relative entropy of entanglement is defined by a variational
formula in a very high dimensional space, whose solution is itself
not easy to do reliably. In addition, the additivity conjecture is
true on a large set in the state space, so unless one has a
specific idea where to look, a random search may well produce
misleading evidence. The second strong point in favor of the
additivity conjecture had been a Theorem by Rains (Theorems 4 and
5 in \cite{Rains}) implying a host of non-trivial additivity
statements. However, our counterexample satisfies the assumptions
of the Rains' Theorem, so that Theorem is, unfortunately, false.

Further specific results in our paper are the formulas for
entanglement of formation and relative entropy of entanglement for
Werner states.

The paper is organized as follows: In Section~II we review the
essential techniques for the investigation of symmetric states and
describe how the partial transposition fit in this context.
Section~II.D presents a zoo of different symmetry groups. Some of
these are used later, others are only presented as briefly, to
illustrate special properties possible in this setup. We hope that
this list will prove useful for finding the right tradeoff between
high symmetry, making an example manageable, and richness of the
symmetric state space, which may be needed to see the phenonmenon
under investigation.  In Section \ref{III} we briefly recapitulate
the definitions of the entanglement of formation and the relative
entropy of entanglement and the additivity problem. In Section~IV
we turn to the entanglement of formation. We show first how the
computation may be simplified using local symmetry. These ideas
are then applied to the basic symmetry groups $\UU$ and $\UUb$,
arriving at an explicit formula in both cases (the results for
$\UUb$ are merely cited here for completeness from work of the
first author with B. Terhal\cite{TerVol}). For the group $\OO$ of
orthogonal symmetries, which unifies and extends these two
examples, we find formulas in large sections of the state space.
Section~V deals with the
relative entropy of entanglement. Again we begin by showing how
the computation is simplified under symmetry. We then present the
counterexample to additivity mentioned in the introduction. Some
possible extensions are mentioned in the concluding remarks.

\section{Symmetries and Partial Transposes}

From the beginning of the theory of entanglement the study of
special subclasses of symmetric states has played an important
role. In this section we give a unified treatment of the
mathematical structure underlying all these studies. For
simplicity we restrict attention to the bipartite finite
dimensional case, although some of the generalizations to more
than two subsystems \cite{tripartite} and infinite dimension are
straightforward. So throughout we will consider a composite
quantum system with Hilbert space $\H=\H_1\otimes\H_2$, with
$\dim\H_i=d_i<\infty$. We denote the  space of states (=density
operators) on $\H$ as $\st(\H)$, or simply by $\st$. The  space of
all separable states (explained in subsection \ref{separabel}) is denoted as $\sep$.

\subsection{Local symmetry groups}\label{hier}
Two states $\rho,\rho'$ are regarded as ``equally entangled'' if
they differ only by a choice of basis in $\H_1$ and $\H_2$ or,
equivalently, if there are unitary operators $U_i$ acting on$H_i$
such that $\rho'=(U_1\otimes U_2)\rho(U_1\otimes U_2)^*$. If in
this equation $\rho'=\rho$, we call $U=(U_1\otimes U_2)$ a (local)
symmetry of the entangled state $\rho$. Clearly, the set of
symmetries forms a closed group of unitary operators on
$\H_1\otimes\H_2$. We will now turn this around, i.e., we fix the
symmetry group and study the set of states left invariant by it.

So from now on, let $G$ be a closed group of unitary operators of
the form $U=(U_1\otimes U_2)$. As a closed subgroup of the unitary
group, $G$ is compact, hence carries a unique measure which is
normalized and invariant under right and left group translation.
Integrals with respect to this {\it Haar measure} will just be
denoted by ``$\int\dU$'', and should be considered as averages
over the group. In particular, when $G$ is a finite group, we have
$\int\dU f(U)=\abs{G}^{-1}\sum_{U\in G}f(U)$. An important
ingredient of our theory is the projection
\begin{equation}\label{twirl}
  \pr(A)=\int\dU\ UAU^* \; ,
\end{equation}
 for any operator $A$ on $\H_1\otimes\H_2$, which is
sometimes referred to as the {\it twirl} operation. It is a
completely positive operator, and is {\it doubly stochastic} in
the sense that it takes density operators to density operators and
the identity operator to itself. Using the invariance of the Haar
measure it is immediately clear that ``$\pr A=A$'' is equivalent
to ``$[U,A]=0$ for all $U\in G$''. The set of all $A$ with this
property is called the {\it commutant} of $G$. We will denote it
by $G'$, which is the standard notation for commutants in the
theory of von Neumann algebras. It will be important later on that
$G'$ is always an algebra (closed under the operator product),
although in general $\pr(AB)\neq(\pr A)(\pr B)$. Computing the
commutant is always the first step in applying our theory.
Typically, one tries to pick a large symmetry group $G$ from the
outset, so the commutant becomes a low dimensional space, spanned
by just a few operators.

Our main interest does not lie in the set $G'$ of $G$-invariant
observables, but dually, in the $G$-invariant density operators
$\rho$ with $\pr\rho=\rho$. As for observables this set is the
projection $\pr\st$ of the full state space under twirling. The
relation between invariant observables and states is contained in
the equation
\begin{equation}\label{pr-star}
  \tr\bigl(\pr(\rho)A\bigr)=\tr\bigl(\rho\pr(A)\bigr) \;,
\end{equation}
which follows easily by substituting $U\mapsto U^*=U^{-1}$ in the
integral (\ref{twirl}), and moving one factor $U$ under the trace.
Due to this equation, we do not need to know the expectations
$\tr(\rho A)$ for all observables $A$ in order to characterize a
$G$-invariant $\rho$, but only for the invariant elements
$\pr(A)\in G'$. Indeed, if we have a linear functional
$f:G'\to\Cx$, which is positive on positive operators, and
normalized to $f(\idty)=1$, that is a {\it state} on the algebra
$G'$ in C*-algebraic terminology, the equation $\tr(\rho
A)=f(\pr(A))$ uniquely defines a $G$-invariant density operator
$\rho$, because $\pr$ preserves positivity and $\pr(\idty)=\idty$.
Under this identification of $G$-invariant density operators and
states on $G'$ it becomes easy to compute the image of a general
density operator $\rho$ under twirling. Using again
equation~(\ref{pr-star}) we find that $\pr\rho$ is determined
simply by computing its expectation values for $A\in G'$, i.e.,
its  {\it restriction} to $G'$.

Let us demonstrate this in the two basic examples of twirling.

\examp1 The group $\UU$ ({\it Werner states}).
We take the Hilbert spaces of Alice and
Bob to be the same ($\H=\H_1\otimes\H_1$), and  choose for $G$ the
group of all unitaries of the form $U\otimes U$, where $U$ is a
unitary on $\H_1$. As an abstract topological group this is the
same as the unitary group on $\H_1$, so the Haar measure on $G$ is
just invariant integration with respect to $U$. It is a well-known
result of group representation theory, going back
to Weyl \cite{Weyl} or further, that the commutant of $G$ is
spanned by the permutation operators of the factors, in this case
the identity $\idty$ and the {\it flip} defined by
$\flip(\phi\otimes\psi)=\psi\otimes\phi$, or in a basis $\ket i$
of $\H_1$, with $\ket{ij}=\ket i\otimes\ket j$:
\begin{equation}\label{flip}
  \flip=\sum_{i,j}\ket{ij}\bra{ji}\;.
\end{equation}
Hence the algebra $G'$ consists of all operators of the form
$A=\alpha\idty+\beta\flip$. As an abstract *-algebra with identity
it is characterized by the relations $\flip^2=\idty$ and
$\flip^*=\flip$. Thus $G$-invariant states are given in terms of the
single parameter $\tr(\rho\flip)$, which ranges from $-1$ to $1$.
Note that an invariant density operator can be written as
$\rho=a\idty+b F$ with suitable $a,b\in\Rl$.  But as we will see,
the parameters $a,b$ are less natural to use, and more dimension
dependent than $\tr(\rho\flip)$.

\examp2 The group $\UUb$ ({\it isotropic states}). Again we take both Hilbert spaces to be
the same, and moreover, we fix some basis in this space. The group
$G$ now consists of all unitaries of the form $U\otimes\overline
U$, where $U$ is a unitary on $\H_1$, and $\overline U$ denotes
the matrix element-wise complex conjugate of $U$ with respect to
the chosen basis. One readily checks that the maximally entangled
vector $\Phi=\sum_i\ket{ii}$ is invariant under such unitaries,
and indeed the commutant is now spanned by $\idty$ and the rank
one operator
\begin{equation}\label{flipt}
  \flipt=\ketbra\Phi\Phi=\sum_{i,j}\ket{ii}\bra{jj}\;.
\end{equation}
 This operator is positive with norm $d=\norm\Phi^2=\dim\H_1$, so
the invariant states are parametrized by the interval $[0,d]$.
These claims can be obtained from the first example by the method
of partial transposition discussed in Subsection~\ref{sec:pt}.

It is perhaps helpful to note that there are not so many functions
$U\mapsto\widetilde{U}$, taking unitaries on $\H_1$ to unitaries
on the same space $\H_1$, such that the operators of the form
$U\otimes\widetilde U$ again form a group. For this it is
necessary that $U\mapsto\widetilde{U}$ is a homomorphism, so, for
example, $\widetilde U=U^*$ does not work. Inner homomorphisms,
i.e., those of the form $\widetilde U=VUV^*$ are equivalent to
Example~1 by a trivial basis change in the second factor, given by
$V$. Similarly, functions differing only by a scalar phase factor
give the same transformations on operators, and should thus be
considered equivalent. Then (up to base changes and phase factors)
all functions $U\mapsto\widetilde{U}$ not equivalent to the
identity are equivalent to Example~2, i.e., the above list is in
some sense complete. However, many interesting examples arise,
when the Hilbert spaces are not of the same dimension, or the
group of operators in the first factor is not the full unitary
group.

Computing $\pr\st$ in Examples 1 and 2 is very simple, because it
is just an interval. We will encounter more complicated cases
below, in most of which, however, the algebra $G'$ is abelian. When
$G'$ has dimension $k$, say, it is then generated by $k$ minimal
projections, which correspond precisely to the extreme points
of $\pr\st$. Therefore the state space is a {\it simplex}
(generalized tetrahedron).

\subsection{How to compute the separable states $\pr\sep$}\label{separabel}

For the study of entanglement of symmetric states it is
fundamental to know which of the states in $\pr\st$ are {\it
separable} or ``classically correlated'' \cite{Wer}, i.e., convex
combinations
\begin{equation}\label{sep}
  \rho=\sum_\alpha\lambda_\alpha\;
        \rho_{1}^{(\alpha)}\otimes\rho_{2}^{(\alpha)}
\end{equation}
of product density operators. We denote this set of states by
$\sep$. Because we assume the group $G$ to consist of local
unitaries, it is clear that for a separable state $\rho$ the
integrand of  $\pr\rho$ consists entirely of separable states,
hence $\pr\rho$ is separable. Hence
$\pr\sep\subset(\sep\cap\pr\st)$. But here we even have equality,
because any state in $(\sep\cap\pr\st)$ is its own projection.
Hence
\begin{equation}\label{prsep}
  \sep\cap\pr\st=\pr\sep\;.
\end{equation}
In order to determine this set, recall that by decomposing
$\rho_{1,2}^{(\alpha)}$ in (\ref{sep}) into pure states, we may
even assume the $\rho_{1,2}^{(\alpha)}$ in (\ref{sep}) to be pure.
If we compute $\pr\rho$ termwise, we find that each
$\rho\in\pr\sep$ is a convex combination of states
$\pr(\sigma_1\otimes\sigma_2)$ with pure
$\sigma_i=\ketbra{\phi_i}{\phi_i}$. Thus we can compute $\pr\sep$
in two stages:
\begin{itemize}
\item Choose a basis in $G'$, consisting, say of $k$ hermitian operators
 $A_\alpha$ and compute the expectations of these operators in arbitrary pure
product states:
 \begin{displaymath}
   a_\alpha=
     \bra{\phi_1\otimes\phi_2}A_\alpha\ket{\phi_1\otimes\phi_2}\;,
\end{displaymath}
this determines the projections $\pr(\sigma_1\otimes\sigma_2)$.
\item Determine the set of real $k$-tuples $(a_1,\ldots,a_k)$
obtained in this way, as the $\phi_i$ range over all normalized
vectors.
\item Compute the convex hull of this set.
\end{itemize}
Two simplifications can be made in this procedure: firstly, we
always have $\idty\in G'$, so by choosing $A_k=\idty$, it suffices
to work with the $(k-1)$-tuples $(a_1,\ldots,a_{k-1})$. Secondly,
the vectors $\phi_1\otimes\phi_2$ and $U(\phi_1\otimes\phi_2)$
with $U\in G$ give the same expectations, so when determining the
range one can make special choices, as long as one vector is
chosen from each orbit of product vectors under $G$.

Let us illustrate this procedure in the two basic examples above:
In Example~1 we only need to compute
\begin{equation}\label{flipD}
  \bra{\phi\otimes\psi}\flip\ket{\phi\otimes\psi}
   =\abs{\braket\phi\psi}^2\;.
\end{equation}
 Clearly, this quantity ranges over the
interval $[0,1]$, and a $UU$-invariant state $\rho$ is separable
iff $\tr(\rho\flip)\geq0$ \cite{Wer}. Similarly, in Example~2:
\begin{equation}\label{fliptD}
  \bra{\phi\otimes\psi}\flipt\ket{\phi\otimes\psi}
   =\abs{\sum_i\phi_i\psi_i}^2
   =\abs{\braket\phi{\overline\psi}}^2\;,
\end{equation}
 which again ranges over the interval
$[0,1]$. Note, however, that the state space in this case is the
interval $[0,d]$. The fact that the two state space intervals
$[-1,1]$ for $\UU$ and $[0,d]$ for $\UUb$ intersect precisely in
the separable subset $[0,1]$ is an instance of the Peres-Horodecki
criterion for separablility, as we now proceed to show.

\subsection{Partial transposition\label{sec:pt}}
The partial transpose of an operator on $\H_1\otimes\H_2$ is
defined in a product basis by transposing only the indices
belonging to the basis of $\H_2$, and not those pertaining $\H_1$.
Equivalently, we can define this operation as
\begin{equation}\label{ptr}
  \ptr(A\otimes B)= A\otimes \Theta(B)\;,
\end{equation}
where $\Theta(B)$ denotes the ordinary matrix transpose of $B$.
This also depends on the choice of basis in $\H_2$, so from now on
we assume a basis of $\H_2$ to be fixed. This equation suffices to
define $\Theta_2$, because all operators on $\H_1\otimes\H_2$ can
be expanded in terms of product operators. The partial transpose
operation has become a standard tool in entanglement theory with
the realization that the partial transpose of a separable density
operator is again positive. This is evident from
Equations~(\ref{sep}) and (\ref{ptr}), and the observation that
the transpose of a positive operator is positive. In $2\otimes2$
and $2\otimes3$ Hilbert space dimensions, this criterion, known as
the Peres-Horodecki criterion, is even sufficient for separability
\cite{Peres}. For all higher dimensions sufficiency fails in
general. States with positive partial transpose (``ppt-states'')
are known not to be {\it distillible}, i.e., even when many copies
of such a state are provided, it is not possible to extract any
highly entangled states by local quantum operations and classical
communication alone.

For special classes of states on higher dimensional Hilbert spaces
the ppt-property may still be sufficient for separability. Pure
states are a case in point, and so are some of the spaces of
symmetric states studied in this paper. Let us check how the
action of a product unitary is modified by partial transposition.
If $U_i,A_i$ are operators on $\H_i$ ($i=1,2$), we find
\begin{eqnarray*}
  \ptr\Big((U_1\otimes U_2&)(&A_1\otimes A_2)
                (U_1^*\otimes U_2^*)\Big)\\
  &=& \Theta_2\Bigl((U_1A_1U_1^*)\otimes(U_2A_2U_2^*)\Bigr)\\
  &=& (U_1A_1U_1^*)\otimes\Bigl(\Theta(U_2^*)\Theta(A_2)\Theta(U_2)\Bigr)\\
     &=&(U_1\otimes\Ubar_2)\Theta_2
         (A_1\otimes A_2)(U_1\otimes\Ubar_2)^*\;.
\end{eqnarray*}
Note that by linearity we can replace in this equation $A_1\otimes
A_2$ by any other operator on $\H_1\otimes\H_2$. This computation
motivates the following definition: For any group $G$ of product
unitaries we denote by $\til G$ the group of unitaries
$U_1\otimes\Ubar_2$, where $U_1\otimes U_2\in G$. For example, for
$G=UU$ of Example~1 we get $\til G=\UUb$, and conversely.

There is a slightly tricky point in this definition, because the
map $U_1\otimes U_2\mapsto U_1\otimes\Ubar_2$ is not well defined:
If we multiply $U_1$ by a phase and $U_2$ with the inverse phase,
the operator $U_1\otimes U_2$ does not change, but
$U_1\otimes\Ubar_2$ picks up twice the phase. What the definition
therefore requires is to take in $\til G$ {\it all} operators
arising in this way. Repeating the ``twiddle'' operation may thus
fail to lead back to $G$, but instead leads to $G$ enlarged by the
group of phases. It is therefore convenient to assume that all
groups under consideration contain the group of phases. We may do
so without loss of generality, since the phases act trivially on
operators anyhow, and hence the twirling projection $\pr$ is
unchanged.

If we integrate the above computation with respect to a
group $G$ of local unitaries, and introduce $\til\pr$ for the
twirling projection associated with $\til G$, we get the
fundamental relation
\begin{equation}\label{tp=pt}
  \ptr\pr=\til\pr\ptr\;.
\end{equation}
Since $\ptr$ is a linear bijection on the space of all operators
on $\H_1\otimes\H_2$, we immediately find the relations between
the ranges of $\pr$ and $\til\pr$:
\begin{equation}\label{tilGprime}
  \ptr(G')=\til G'\;,
\end{equation}
i.e., the operators invariant under $\til G$ are precisely the
partial transposes of those invariant under $G$. This has a
surprising consequence: taking the partial transposes of an
algebra of operators in general has little chance of producing
again an algebra of operators, since $\ptr$ is definitely not a
homomorphism. That is, in general one would not expect that the
operator product of two partial transposes is again the partial
transpose of an element of the original algebra. If the algebra
arises as the commutant of a group of {\it local} unitaries, however,
we get again a commutant, hence an algebra.

The first application of Equation~(\ref{tilGprime}) is the
computation of the commutant in Example~2: With $G=UU$ we find the
partial transposes of the operators in $G'$, i.e., the operators
$\ptr(\alpha\idty+\beta\flip)=\alpha\idty+\beta\flipt$, since
$\ptr(\flip)=\flipt$.

Another application is the determination of the set of ppt-states.
One might think that a special form for $\rho$, entailed by its
$G$-invariance, is not necessarily helpful for getting spectral
information about $\ptr\rho$. However, since $\ptr G'$ is an
algebra, and often enough an abelian one, $\ptr\rho$ is, in fact,
easily diagonalized.

A good way to represent this connection is to draw the state
spaces of $G$ and $\til G$ (i.e., $\pr\st$ and $\til\pr\st$) in
the same diagram. Since in general $G'$ and $\til G'$ need not
intersect except in the multiples of the identity (see Examples 1
and 2), the projected state spaces $\pr\st$ and $\til\pr\st$ in
general have only the trace-state in common. Hence they don't fit
naturally in the same diagram. However, the partial transposes of
$\til\pr\st$ lie in $G'$, more precisely in the hyperplane of
hermitian elements with trace $1$. The same hyperplane contains
$\pr\st$. In the pair of Examples 1 and 2, we get Figure~1.
\begin{figure}
\begin{center}
\epsfxsize=8.5cm \epsffile{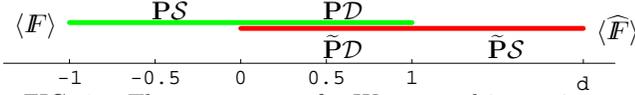}
\caption{ The state spaces for Werner and isotropic states
are just intervals. Drawn together in one diagram the intersection
gives us the space of PPT-states, which is in this case equivalent to the separable
space.  }
\label{fig1}
\end{center}
\end{figure}

Note that by exchanging the roles of $G$ and $\til G$, we get
exactly the same diagram, up to maybe an affine transformation due
to a different choice of coordinates: the two diagrams are simply
related by taking partial transposes. When $G$ and $\til G$ are
swapped in this way, the picture of $\pr\sep$ remains correct:
since $\ptr\sep=\sep$, it suffices to compute the projection of
the separable subset for $G$. By definition, the intersection of
$\pr\st$ and $\ptr\til\pr\st$ is the convex set of $G$-invariant
ppt-states. It always contains $\pr\sep$, but this inclusion may
be strict. In the simple case of Figure~1
$\pr\sep=\pr\st\cap\ptr\til\pr\st$, which is the same as saying
that the Peres-Horodecki criterion is valid for states invariant
under either $G$ or $\til G$.

\subsection{Further examples of symmetry groups}

\examp3 Orthogonal groups: $G=\OO$.
 The two basic examples can be combined into one by taking the
{\it intersection} of the two groups: $G=UU\cap\UUb$ this is the
same as the subgroup of unitaries $U\otimes U$ such that
$\overline U=U$, i.e., such that $U$ is a real orthogonal matrix.
Clearly, both the $UU$-invariant states and the $\UUb$-invariant
states will be $G$-invariant, so we know that $G'$ is at least%
\footnote{In general, the commutant $(G\cap H)'$ may be properly
larger than the algebra $G'\vee H'$ generated by $G'$ and $H'$.
The equation $({\cal A}\cap{\cal B})'={\cal A}'\vee{\cal B}'$ is
valid only for algebras, and follows readily from the equation
$({\cal A}'\vee{\cal B}')'=({\cal A}''\cap{\cal B}'')$, and the
bicommutant theorem \cite{bicommutant}, which characterizes $M''$
as the algebra generated by $M$. However, the algebras $G''$ and
$H''$ may have an intersection, which is properly larger than the
algebra generated by their intersection. For example, for any
irreducible represented group $G''$ is the algebra of all
operators, but two such groups may intersect just in the identity.
Hence some caution has to be exercised when computing $(G\cap H)'$
for general groups.}%
the algebra generated by $UU'$ and $\UUb'$, i.e., it contains
$\idty,\flip$, and $\flipt$. Since
$\flip\flipt=\flipt\flip=\flipt$, the linear span of these three
is already an algebra, and is spanned by the minimal projections
\begin{eqnarray}
  p_0&=&\frac13 \flipt\\
  p_1&=&\frac12 (\idty-\flip)\\
  p_2&=&\frac12 (\idty-\flip)-\frac13 \flipt\;,
\end{eqnarray}
which corresponds precisely to the decomposition of a general
$3\times3$-matrix into multiple of the identity, antisymmetric
part, and symmetric traceless part. This decomposition of tensor
operators with respect to the orthogonal group is well known, so
we have identified $G'$.

\begin{figure}
\begin{center}
\epsfxsize=8.5cm \epsffile{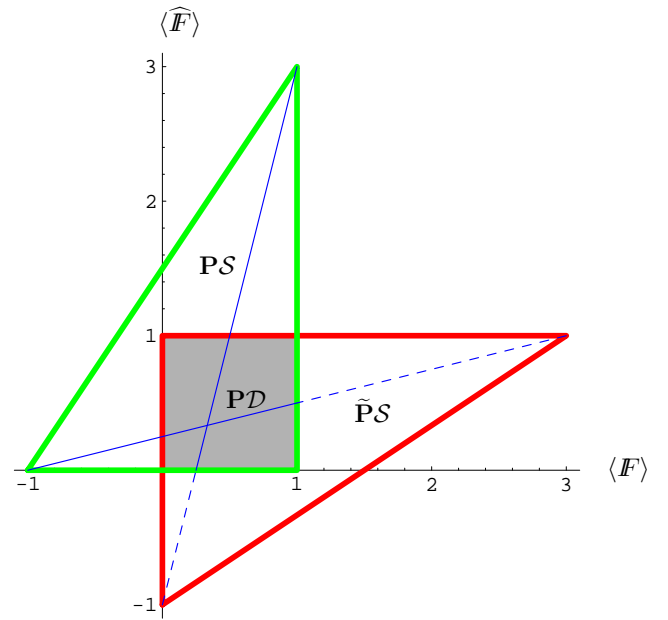} \caption{State spaces for
$OO$ and $O \bar O$ invariant states plotted for $d=3$. The $UU$
and $\UUb$ invariant states are drawn as thin lines.} \label{fig2}
\end{center}
\end{figure}

The extremal $G$-invariant states corresponding to these three
minimal projections are plotted in Figure~\ref{fig2} in a coordinate
system whose axes represent the expectations of $\flip$ and
$\flipt$, respectively. The plane of this drawing should be
considered as the hermitian $G$-invariant operators of trace one.
This plane is mapped into itself by partial transposition (since
$G=\til G$), and the coordinates are chosen such that partial
transposition is simply the reflection along the main diagonal.
The intersection of $\pr\st$ and $\til\pr\st$ is the square
$[0,1]\times[0,1]$. Is the Peres-Horodecki criterion valid for
these states? All we have to do to check this is to try to get
some pure product states, whose expectations of $\flip$ and
$\flipt$ fall on the corners of this square. For a product vector
$\phi\otimes\psi$ we get the pair of expectations
\begin{displaymath}
  \Bigl(\abs{\braket\phi\psi}^2      \;,\
  \abs{\braket\phi{\overline\psi}}^2 \Bigr)\;.
\end{displaymath}
Here $\overline\psi$ denotes the complex conjugate of $\psi$ in a
basis in which the representation is real. Now the point $(1,1)$
in the square is obtained, whenever $\phi=\psi$ is real, the point
$(0,0)$ is obtained when $\phi$ and $\psi$ are real and
orthogonal, and the point $(1,0)$ is obtained when $\psi=\phi$,
and $\braket\phi{\overline\phi}=0$, for example
$\phi=(1,i,0)/\sqrt2$. Symmetrically we get $(0,1)$ with the same
$\phi$ and $\psi=\overline\phi$. Hence all four corners are in
$\pr\sep$, and as this is a convex set we must have
$\pr\sep=\pr\st\cap\ptr(\til\pr\st)$.

\examp4 $\SU2$-representations.
 A class of examples, in which arbitrary dimensions of
$\H_1$ and $\H_2$ can occur is the following. Let
$u\mapsto\surep^j_u$ denote the spin $j$ irreducible
representation of $\SU2$. Then we can take
\begin{equation}\label{Gex3}
  G=\Set\big{\surep^{j_1}_u\otimes\surep^{j_2}_u}{u\in\SU2}\;,
\end{equation}
where $(2j_k+1)$ is the dimension of $\H_k$ ($k=1,2$). Since the
$j_k$ also take half-integer values, these dimensions can be any
natural number $\geq1$. It is known from just about any quantum
mechanics course (under the key word ``addition of angular
momenta'') that the tensor product representation
$\surep^{j_1}\otimes\surep^{j_2}$ is decomposed into the direct
sum of the irreducible representations $\surep^s$ with
$s=\abs{j_1-j_2},\abs{j_1-j_2}+1,\ldots(j_1+j_2)$, each of these
representations appearing with multiplicity $1$. Therefore, the
commutant of $G$ is spanned by the projections onto these
subspaces, and is an abelian algebra.

Note that since the spin-$1$ representation of $\SU2$ is the
orthogonal group in $3$ dimensions, the case $j_1=j_2=1$
corresponds precisely to the previous example with $d=3$. We have
no general expression for the separable subsets, nor even for the
partially transposed sets in these examples. We believe, however,
that this class of examples deserves further investigation.

\examp5 Bell diagonal states.
 In this example we show that the group $G$ can also be abelian,
and we make contact with a well investigated structure of the two
qubit system. So let  $\H_1=\H_2=\Cx^2$, and  let $\sigma_k$,
$k=1,2,3$ be the Pauli matrices, and $\sigma_0=\idty$. Then the
set
\begin{equation}\label{PauliG}
  G=\{\idty,\; -\sigma_1\otimes\sigma_1,\; -\sigma_2\otimes\sigma_2,\;
             -\sigma_3\otimes\sigma_3\}
\end{equation}
forms a group, which is isomorphic to the Klein $4$-group, and
abelian ($G\subset G'$). It is even maximally abelian, i.e, the
algebra $G''$ generated by $G$ is equal to, and not just contained
in $G'$. The minimal projections in $G'$ are
$\ketbra{\Psi_k}{\Psi_k}$, $k=0,1,2,3$, where the $\Psi_k$ are the
magical {\it Bell Basis} \cite{EoF,WV}:
$\Psi_0=(\ket{11}+\ket{22})/\sqrt2$, and
$\Psi_k=i(\idty\otimes\sigma_k)\Psi_0$ for $k=1,2,3$. In this
basis the group elements and their negatives are the diagonal
operators with diagonal elements $\pm1$, of which an even number
are $-1$. Hence the $G$-invariant states are the tetrahedron of
density operators which are diagonal in Bell Basis.

The partial transpose is easy to compute: only $\sigma_2$ changes
sign under transposition. Hence if we draw the state space in a
coordinate system, whose three axes are the expectations of the
group elements $-\sigma_k\otimes\sigma_k$ ($k=1,2,3$), the Bell
states are the corners $(1,1,1)$, $(1,-1,-1)$, $(-1,-1,1)$, and
$(-1,1,-1)$ of the unit cube, from which their partial transposes
are obtained by mirror reflection $x_2\mapsto-x_2$. That is, the
partially transposed states occupy the remaining four corners of
the unit cube. The ppt-subset, which is equal to the separable
subset since we are in $2\otimes2$ dimensions, is hence the
intersection of two tetrahedra, and is easily seen to be an
octahedron.

\begin{figure}
\begin{center}
\epsfxsize=8.5cm \epsffile{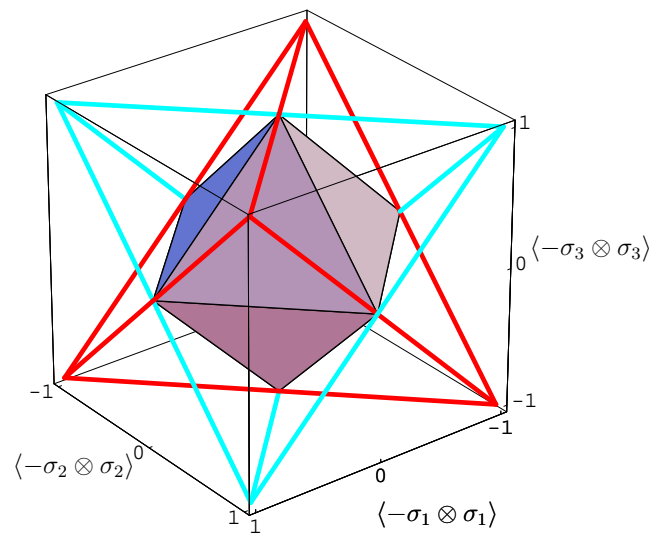}
\caption{State spaces for Bell diagonal states. }
\label{fig3}
\end{center}
\end{figure}

\examp6 Finite Weyl Systems.
 In the examples so far the groups $G$ and $\til G$ were
isomorphic or even equal. In this example, which extends the
previous one, we see that the two groups and their commutants can
be very different.

We let $d$ be an integer, and introduce on $\Cx^d$
the {\it Weyl operators}, given by
\begin{equation}
 W(x,y)\ket z=\omega^{xz}\ket{z-y}\;,
\end{equation}
 where $\omega=\exp(2\pi i/p)$. These are unitary, and satisfy the
 ``Weyl relations''
\begin{equation}\label{Weylrel}
  W(x_1,y_1)W(x_2,y_2)=\omega^{-x_1y_2}W(x_1+x_2,y_1+y_2)\;.
\end{equation}
Hence these operators, together with the $p^{\rm th}$ roots of
unity form a group. On $\Cx^d\otimes\Cx^d$ we introduce the
operators  $W(x_1,y_1,x_2,y_2)\equiv W(x_1,y_1)\otimes
W(x_2,y_2)$, and take
\begin{equation}\label{GWeyl}
  G=\Set\big{\omega^zW(x,y,x,y)\;}{x,y,z=0,\ldots,d-1}\;.
\end{equation}
The commutant is readily computed from the Weyl relations to be
\begin{equation}\label{GWeyl}
  G'={\rm span}\Set\big{W(x,y,-x,-y)\;}{x,y=0,\ldots,d-1}\;.
\end{equation}
The Weyl operators in $G'$ satisfy Weyl relations with $\omega$
replaced by $\omega^2$. If $d$ is odd, such relations are
equivalent to the Weyl relations (\ref{Weylrel}) for a
$d$-dimensional system, and hence $G'$ is isomorphic to the
$d\times d$-matrices.

On the other hand, complex conjugation of $W(x,y)$ just inverts
the sign of $x$, so $\til G$ contains the Weyl operators
$W(x,y,-x,y)$. But this time, rather than getting twice the Weyl
phase, the phases cancel, and $\til G$ is {\it abelian}. One also
verifies that
\begin{equation}\label{GWeyl}
  \til G'={\rm span}\Set\big{W(x,y,-x,y)\;}{x,y=0,\ldots,d-1}\;.
\end{equation}
is spanned by $\til G$, so this algebra is even maximally abelian:
it contains $d^2$ one-dimensional projections, which thus form the
extreme points of $\til\pr\st$. Hence we get the following
picture: the set $\pr\st$ of $G$-invariant states is isomorphic to
the space of $d\times d$-density operators, and the $G$-invariant
operators with positive partial transpose are a simplex spanned by
$9$ extreme points, which are mapped into each other by the action
of a $d\times d$ Weyl system. The intersection
$\pr\st\cap\ptr(\til\pr\st)$ is a rather complicated object. We do
not know yet whether it differs from $\pr\sep$.

\examp7 Tensor products.
 Additivity problems for entanglement (see Section~\ref{sec:add}
for a brief survey) concern tensor products of bipartite states,
which are taken in such a way as to preserve the splitting between
Alice and Bob. Thus in the simplest case we have four subsystems,
described in Hilbert spaces $\H_i,\K_i$, $i=1,2$, such that
systems $\H_1$ and $\K_1$ belong to Alice, systems $\H_2$ and
$\K_2$ belong to Bob, and such that the systems in $\H_i$  are
prepared together according to a density matrix $\rho$ on
$\H_1\otimes\H_2$ and, similarly, the remaining systems are
prepared according to $\sigma$, a density operator  on
$\K_1\otimes\K_2$. We wish to study the entanglement properties of
$\rho \otimes\sigma$, when both these density matrices are assumed
to be invariant under suitable groups of local unitaries.

Let us denote by $G$ (resp. $H$) the group of local unitaries on
$\H_1\otimes\H_2$ (resp. by $\K_1\otimes\K_2$), and assume $\rho$
and $\sigma$ to be invariant under the respective group. Then,
clearly, $\rho\otimes\sigma$ is invariant under all unitaries
$U_1\otimes U_2\otimes V_1\otimes V_2$, where
 $U_1\otimes U_2\in G$ and $V_1\otimes V_2\in H$. These again form
a group of local unitaries, denoted by  $G\otimes H$, where
``local'' is understood in the sense of the Alice$-$Bob splitting
of the system, i.e., the unitary $U_1\otimes V_1$ acts on Alice's
side and $U_2\otimes V_2$ on Bob's. In this sense the product
state is invariant under the group $G\otimes H$ of local
unitaries, and we can apply the methods developed below to compute
various entanglement measures for it.

Computing the commutant $(G\otimes H)'$ is easy, because we do not
have to look at the Alice$-$Bob splitting of the Hilbert space. In
fact, we can invoke the ``Commutation Theorem'' for von Neumann
algebras to get
\begin{equation}\label{GHprime}
  (G\otimes H)'=G'\otimes H'\;,
\end{equation}
where the notation on the right hand side is the tensor product of
algebras, i.e., this is the set of all linear combinations of
elements of the form $A\otimes B$ where $A\in G'$ acts on the
first two and $B\in H'$ acts on the second two factors of
$\H_1\otimes\H_2\otimes\K_1\otimes\K_2$. In particular, if $G'$
and $H'$ are abelian, so is $G'\otimes H'$, and we can readily
compute the minimal projections, which correspond to the extremal
invariant states: if $p_\alpha$ are the minimal projections of
$G'$ and $q_\beta$ are those of $H'$, then the minimal projections
of $G'\otimes H'$ are all $p_\alpha\otimes q_\beta$.

Partial transposition also behaves naturally with respect to
tensor products, which implies that
 $(G\otimes H)\til{{\ }}=\til G\otimes\til H$,
and allows us to compute in a simple way the $(G\otimes
H)$-invariant states with positive partial transpose from the
corresponding data of $G$ and $H$. However, for the determination
of $\pr\sep$ no such shortcut exists.

\begin{figure}
\begin{center}
\epsfxsize=8.5cm \epsffile{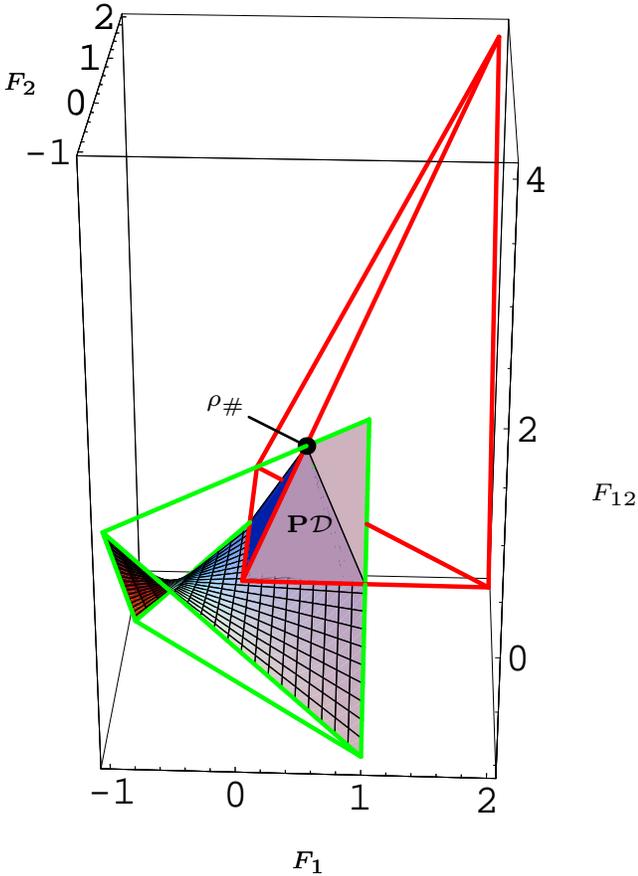} \caption{ State space for
$UUVV$-invariant states plotted for dimension $d=3$. }
\label{fig4}
\end{center}
\end{figure}

We illustrate this in the example, which we will also use for the
counterexample to additivity of the relative entropy of
entanglement announced in the Introduction. For this we take
$G=H=UU$, with a one-particle space $\H_1=\H_2=\K_1=\K_2=\Cx^d$,
for any dimension $d<\infty$. The extreme points of the state
space of $G'$ are given by the normalized projections
\begin{equation}\label{projpm}
  \rho_\pm=\frac1{d(d\pm 1)}(\idty\pm\flip)\;.
\end{equation}
Hence the state space of the abelian algebra $(G'\otimes H')$ is
spanned by the four states $\rho_{s_1}\otimes \rho_{s_2}$,
$s_1,s_2=\pm$ and is a tetrahedron. A convenient coordinate system
is given by the expectations of the three operators
\begin{eqnarray}\label{flip1212}
  F_1   &=&\flip\otimes\idty\\
  F_2   &=&\idty\otimes\flip\\
  F_{12}&=&\flip\otimes\flip\;.
\end{eqnarray}
The four extreme points are then on the edges of the unit cube:
$\rho_{s_1}\otimes \rho_{s_2}$ has expectation triple
$(s_1,s_2,s_1s_2)$. This is drawn in Figure~\ref{fig4}.

The extreme points are special instances of product states: when
$\rho,\sigma$ are $UU$-invariant states with flip expectations
$f_1$ and $f_2$, respectively, the product state
$\rho\otimes\sigma$ has coordinates $(f_1,f_2,f_1f_2)$. Hence the
manifold of product states is embedded in the state space as a
piece of hyperboloid. Partial transposition turns the flip
operators (\ref{flip1212}) into their counterparts using $\flipt$
instead of $\flip$. Hence the operators with positive partial
transposes are represented in the diagram by a tetrahedron with
vertices $(0,0,0)$, $(0,d,0)$, $(d,0,0)$, and $(d,d,d^2)$. The
intersection, i.e., the set of states with trace equal to one and positive partial
transpose (represented in Figure~\ref{fig4} as a solid) is a
polytope with the five extreme points  $(0,0,0)$, $(0,1,0)$,
$(1,0,0)$,  $(1,1,1)$, and, on the line connecting the origin to
the point $(d,d,d^2)$, the point $(1/d,1/d,1)$. The density
operator corresponding to this last point is
\begin{equation}\label{rhostar}
  \rho_\#=\frac{d+1}{2d}\rho_+\otimes\rho_+
         +\frac{d-1}{2d}\rho_-\otimes\rho_-\;.
\end{equation}
It turns out that $\rho_\#$ is separable: Let
$\Phi=d^{-1/2}\sum_k\ket{kk}$ be a maximally entangled vector, and
consider a pure state with vector
$\Psi=\Phi_{\text{Alice}}\otimes\Phi_{\text{Bob}}$. Note that this
is a tensor product with respect to the splitting Alice$-$Bob,
i.e., $13\vert24$ rather than the splitting between pair~1 and
pair~2, i.e., $12\vert34$. We claim that upon twirling this pure
state becomes $\rho_\#$. For this we only need to evaluate the
expectations of the three operators (\ref{flip1212}), and compare
with those of $\rho_\#$. Clearly, $\Psi$ is a symmetric product
(Bose-) vector with respect to the  total flip $F_{12}$, hence
this operator has expectation $1$. The expectations of $F_1$ and
$F_2$ are equal to
\begin{eqnarray*}
  \braket\Psi{F_1\Psi}
    &=&\frac1{d^2}\sum
      \bra{ijij}(\flip\otimes\idty)\ket{k\ell k\ell}\\
    &=&\frac1{d^2}\sum_{i,j,k,\ell}\braket{ijij}{\ell kk\ell}
     =\frac1d\;.
\end{eqnarray*}
Since the other four extreme points are separable as tensor
products of separable states, we conclude that all ppt-states are
separable in this example, so the solid in Figure~\ref{fig4}
also represents the separable subset.

\examp8 Tripartite symmetry: $U\otimes(U\otimes U)$.

The idea of symmetry can also be used to study multi-partite
entanglement. A natural choice of symmetry group is the group of
all unitaries of the form $U\otimes U\otimes U$. The resulting
five dimensional state space has been studied in great detail in
\cite{tripartite}. This study also has a bipartite chapter, where
this group is considered as a group of local unitaries
$U\otimes(U\otimes U)$ in the sense of the present paper. The set
of separable states is strictly smaller than the set of states
with positive partial transposes. However, if we enlarge the group
to include the unitary $\idty\otimes\flip$, the two once again
coincide, forming a tetrahedron.

\section{Entanglement measures and additivity}\label{III}

\subsection{Entanglement of Formation and the convex
   hull construction for functions\label{sec:eofdef}}

The entanglement of a pure state is well described
by the von Neumann entropy of its restricted density operator.
Thus for a pure state $\rho=\ketbra\Psi\Psi$ such that $\Psi$ is
expressed in Schmidt form as $\Psi=\sum_k \sqrt{c_k}e_k\otimes
e_k'$, we have
\begin{eqnarray}\label{Epure}
  E(\rho)&=&\sum_k\eta(c_k)\quad\text{with}\\
  \eta(t)&=&-t\log(t)\;.\label{eta}
\end{eqnarray}
The {\it entanglement of formation} is a specific extension of
this function to mixed states. The extension method is a general
one, known as the {\it convex hull} construction for functions,
and since we will need this construction for stating our main
result, we will briefly review it.

So let $K$ be a compact convex set, let $M\subset K$ be an
arbitrary subset, and let $f:M\to\Rplus$. We then define a
function $\co f:K\to\Rplus$ by
\begin{equation}\label{co1}
   \co f(x)=\inf\Set\Big{\sum_i\lambda_if(s_i)}{s_i\in M,\
                         \sum_i\lambda_is_i=x}\;,
\end{equation}
where the infimum is over all convex combinations with
$\lambda_i\geq0$, $\sum_i\lambda_i=1$, and by convention the
infimum over an empty set is $+\infty$. The
name ``convex hull'' of this function is due to the property that
$\co f$ is the largest convex function, which is $\leq f$ at all
points, where $f$ is defined. Another way of putting this is to
say that the ``supergraph'' of $\co f$, i.e., $\Set\big{(x,r)\in
K\times\Rl}{r\geq\co f(x)}$, is the convex hull (as a subset of
$K\times\Rl$) of $\Set\big{(x,r)\in K\times\Rl}{x\in M,\  r\geq
f(x)}$.

In this notation, the usual definition \cite{EoF} of entanglement of
formation is then
\begin{equation}\label{EoFco}
  \EoF(\rho)=(\co E)(\rho)\;,
\end{equation}
where on the right hand side $E$ is understood as the
function~(\ref{Epure}) defined only on the submanifold
$M\subset\st(\H_1\otimes\H_2)$ of pure states.

\subsection{Relative Entropy of Entanglement}
Another measure of entanglement, originally proposed in \cite{ERE}
is based on the idea that entanglement should be zero for
separable density operators (see Equation~(\ref{sep})), and should
increase as we move away from $\sep$. Such a function might be
viewed as measuring some kind of distance of the state to the set
$\sep$ of separable states. If one takes this idea literally, and
uses the relative entropy \cite{Petz}
\begin{equation}\label{relent}
  \SS(\rho,\sigma)=\tr\rho(\log\rho-\log\sigma)
\end{equation}
to measure the ``distance'', one arrives at the {\it relative
entropy of entanglement}
\begin{equation}\label{ere}
  \ERE(\rho)=\inf\Set\big{\SS(\rho,\sigma)}{\sigma\in\sep}\;.
\end{equation}
Initially, other distance functions have also been used to define
measures of entanglement. However, the one based on the relative
entropy is the only proposal, which coincides on pure states with
the ``canonical'' choice described in Equation~(\ref{Epure}).
Since $\ERE$ is easily shown to be convex, it must be smaller than
the largest convex function with this property, namely $\EoF$.
Another reason to prefer relative entropy over other distance-like
functionals is that it has good additivity properties. The hope
that $\ERE$ might be additive was borne out by initial
explorations, and has become a folk conjecture in the field.
However, we will give a counterexample below.

\subsection{Additivity}\label{sec:add}
A key problem in the current discussion of entanglement measures
is the question, which of these are ``additive'' in the following
sense: if $\rho , \sigma$ are bipartite states on the Hilbert
spaces $\H_1\otimes\H_2$ and $\K_1\otimes\K_2$, then
$\rho\otimes\sigma$ is a state on
$\H_1\otimes\H_2\otimes\K_1\otimes\K_2$. After sorting the factors
in this tensor product into spaces $\H_1 , \K_1$ belonging to
Alice and $\H_2, \K_2$ belonging to Bob, we can consider
$\rho\otimes\sigma$ as a bipartite state on
$(\H_1\otimes\K_1)\otimes(\H_2\otimes\K_2)$. This corresponds
precisely to the situation of a source distributing particles to
Alice and Bob, $\rho\otimes\sigma$, and similar larger tensor
products, being interpreted as the state obtained by letting Alice
and Bob {\it collect} their respective particles. Additivity of an
entanglement measure $E$ is then the equation
\begin{equation}\label{addi}
  E(\rho\otimes\sigma)=E(\rho)+E(\sigma)\;.
\end{equation}
We speak of {\it subadditivity} if ``$\leq$'' holds instead of
equality here. Both $\ERE$ and $\EoF$ are defined as infima, and
for a product we can insert tensor products of convex
decompositions or closest separable points into these infima, and
use the additivity properties of entropy to get subadditivity in
both cases. It is the converse inequality, which presents all the
difficulties, i.e., the statement that in these minimization
problems the tensor product solutions (and not some entangled
options) are  already the best.

Additivity of an entanglement functional is a strong expression of
the {\it resource character} of entanglement. According to an
additive functional, sharing two particles from the same preparing
device is exactly ``twice as useful'' to Alice and Bob as having
just one. Here preparing two pairs means preparing {\it
independent} pairs, expressed by the tensor product in
(\ref{addi}). It is interesting to investigate the influence of
correlations and entanglement between the different pairs. On the
one hand, Alice and Bob might not be aware of such correlations,
and use the pairs as if they were independent. On the other hand,
they might make use of the exact form of the state, including all
correlations. Is the second possibility always preferable?
Entanglement functionals answering this question with ``yes'' have
a property stronger than additivity, called {\it strong
superadditivity}. It is written as \begin{equation}\label{ssadd}
  E(\rho)\geq E(\rho_\H)+E(\rho_\K)\;,
\end{equation}
where $\rho$ is a density operator for two pairs (four particles
altogether), and $\rho_\H$ and $\rho_\K$ are the restrictions to the
first and second pair. An entanglement functional satisfying this
as well as subadditivity is clearly additive. Since additivity is
already difficult to decide, it is clear that strong
superadditivity is not known for any of the standard measures of
entanglement.

One case of strong superadditivity is satisfied both for $\EoF$
and $\ERE$, and we establish this property here in order to get a
more focused search for counterexamples later on: We claim that
(\ref{ssadd}) holds, whenever $\rho_\K$ is separable, in which
case, of course, the second term on the right vanishes (as a
special case of additivity, when $\rho_\K$ is even a product, this
was noted recently in \cite{BeNar}). We will show this by
establishing another property, called {\it monotonicity}: for both
$E=\EoF$ and $E=\ERE$, we claim
\begin{equation}\label{monotone}
  E(\rho)\geq  E(\rho_\H)\;.
\end{equation}

Monotonicity for $\ERE$ follows readily from a similar property of
the relative entropy: if $\rho_\H,\sigma_\H$ denote the
restrictions of states $\rho,\sigma$ to the same subsystem, then
$\SS(\rho_\H,\sigma_\H)\leq\SS(\rho,\sigma)$. But if $\sigma$ is
separable in (\ref{ere}), then so is its restriction $\sigma_\H$.
The infimum over {\it all} separable states on $\H_1\otimes\H_2$
is still smaller, hence monotonicity holds.

Monotonicity for $\EoF$ is similar: We may do the reduction in
stages, i.e., first reduce Alice's and then Bob's system, and
because  $\EoF$ symmetric with respect to the exchange of Alice
and Bob, it suffices to consider the case of a reduction on only
one side, i.e., the restriction from $\H_1\otimes(\H_2 \otimes
\K_2)$ to $\H_1 \otimes \H_2$.

Let $\rho$ be a state on $\H_1\otimes(\H_2\otimes\K_2)$ and
$\rho'$ its restriction to $\H_1\otimes\H_2$.

Consider the states $s_i$ on the larger space appearing in the
minimizing convex decomposition of $\rho$, and let $s_i'$ denote
their restrictions to $\H_1 \otimes \H_2$. Of course, both $s_i$
and $s_i'$ have the same restriction to the first factor $ \H_1$.
Hence
\begin{equation}\label{EoFmon}
  \EoF(\rho)=\sum_i\lambda_if(s_i')\;,
\end{equation} where $f(\sigma)$ denotes the von Neumann entropy
of the restriction of a state $\sigma$ to $\H_1$, and
$\sum_i\lambda_is_i'=\rho'$. Because the entropy of the
restriction is a concave function, the value of the sum
(\ref{EoFmon}) can be made smaller by replacing each $s_i'$ with a
decomposition into pure states on $\H_1\otimes\H_2$. Minimizing
over all such decompositions of $\rho'$ yields $\EoF(\rho')$,
which is hence smaller than $\EoF(\rho)$.

\section{Entanglement of formation}
\subsection{Simplified computation}

Our method for computing the entanglement of formation can also be
explained in the general setting of the convex hull construction
in Subsection~\ref{sec:eofdef}, and this is perhaps the best way to
see the geometrical content. So in an addition to a subset
$M\subset K$ of a compact convex set and a function
$f:M\to\Rplus$, consider a compact group $G$ of symmetries acting
on $K$ by transformations $\alpha_U:K\to K$, which preserve convex
combinations. We also assume that $\alpha_UM\subset M$, and
$f(\alpha_Us)=f(s)$ for $s\in M$. All this is readily verified for
$\alpha_U(A)=UAU^*$ and $f$ the entanglement defined on the subset
$M\subset K$ of pure bipartite states. Our task is to compute $\co
f(x)$ for all $G$-invariant $x\in K$, i.e., those with
$\alpha_U(x)=x$ for all $U\in G$.

Since the integral with respect to the Haar measure is itself a
convex combination, we can define, as before, the projection
$\pr:K\to K$ by $\pr x=\int\dU\alpha_U(x)$. The set of projected
points $\pr x$ will be denoted by $\pr K$. Usually, this will be a
much lower dimensional object than $K$, so we will try to reduce
the computation of the infimum (\ref{co1}), which involves a
variation over all convex decompositions of $x$ in the high
dimensional set $K$ to a computation, which can be done entirely
in $\pr K$. To this end, we define the function $\eps:\pr
K\to\Rplus$ by
\begin{equation}\label{eps}
  \eps(x)=\inf\Set\big{f(s)}{s\in M,\ \pr s=x}\;,
\end{equation}
again with the convention that the infimum over the empty set is
$+\infty$. Then the main result of this subsection is that, for
$x\in\pr K$,
\begin{equation}\label{cofsym}
  \co f(x)=\co\;\eps(x)\;,
\end{equation}
where the convex hull on the left is defined by (\ref{co1}), but
the convex hull on the right is now to be computed in the convex
subset $\pr K$.

We thus arrive at the following recipe for computing the
entanglement of formation of $G$-invariant states:
\begin{itemize}
\item Find, for every state $\rho\in\pr\st$, the set $M_\rho$ of
pure states $\sigma$ such that $\pr\sigma=\rho$.
\item Compute
\begin{equation}\label{epsEoF}
  \eps(\rho):=\inf\Set\big{E(\sigma)}{\sigma\in M_\rho}\;.
\end{equation}
\item For later use try to get a good understanding of the pure
states achieving this minimum.
\item Compute the convex hull of the function (\ref{epsEoF}).
\end{itemize}
The following simplifications are sometimes possible: first of
all, all pure states in an orbit of $G$ give the same value of
$E$, hence we may replace $M_\rho$ by a suitably parametrized
subset containing at least one element from every orbit. At this
stage it is sometimes already possible to discard further states,
in favour of others ``obviously'' giving a smaller value of $E$.
The final stage is sometimes carried out by showing that the
function $\eps$ is convex to begin with, but, as we will see, this
is not always the case.

The remainder of this subsection is devoted to the proof of
Equation~(\ref{cofsym}). We will proceed by showing that both
sides are equal to
\begin{equation}\label{ZZ}
  Z=\inf\Set\Big{\sum_i\lambda_if(s_i)}{s_i\in M,\
                         \sum_i\lambda_i\pr s_i=x}\;.
\end{equation}
Indeed, the only difference between (\ref{ZZ}) and (\ref{co1}) is
that in (\ref{ZZ}) a weaker condition is demanded on the $s_i$.
Hence more $s_i$ are admissible, and this infimum is smaller,
$Z\leq\co f(x)$. On the other hand, if $s_i$ satisfying the
constraint for $Z$ are given, inserting the definition of $\pr$
produces a convex combination giving $x$, namely the combination
of the states $\alpha_U(s_i)$, labeled by the pair $(i,U)$, and
weighted with $\sum_i\lambda_i\int\dU$. This convex combination is
admissible for the infimum defining $\co f$, and gives the value
$\sum_i\lambda_i\int\dU f(\alpha_U(s_i))=\sum_i\lambda_i \int\dU
f(s_i)=\sum_i\lambda_if(s_i)$, where we have used the invariance
property of $f$ and the normalization of the Haar measure. Hence
all numbers arising in the infimum (\ref{ZZ}) also appear in the
infimum (\ref{co1}), which proves that $Z\leq\co f(x)$, hence
$Z=\co f(x)$. In order to prove the equality $Z=\co\;\eps(x)$ just
note that in the infimum (\ref{ZZ}) the constraint is only in
terms of $\pr s_i$, whereas the functional to be minimized
involves $f(s_i)$. Therefore we can compute the infimum (\ref{ZZ})
in stages, by first fixing all $\pr s_i$ and minimizing each
$f(s_i)$ under this constraint, which amounts to replacing $f$ by
$\eps$, and then varying over the $\pr s_i$, which is the infimum
defining $\co\;\eps$. Hence $\co\;\eps(x)=Z=\co f(x)$.

\subsection{Extending the computation to some non-symmetric
states}\label{sec:ext}
It is a basic feature of the convex hull that whenever the infimum
in (\ref{co1}) is found at a non-trivial convex combination, there
is a ``flat piece'' in the graph of $\co f$, i.e., $\co f$ is also
known on the convex hull of the minimizing $s_i$
\cite{Uhlmann}.
The geometrical
meaning of this elementary observation is immediately clear from
low dimensional pictures. It is also easy to prove in general:

Suppose that $\sum_i\lambda_is_i=x$ is a convex decomposition of
$x$ (with $\lambda_i>0$) minimizing $\sum_i\lambda_if(s_i)$, and
let $x'=\sum_i\lambda'_is_i$ be another convex combination of the
same points $s_i$. We claim that this convex combination solves
the minimization problem for $\co f(x')$, i.e.,
\begin{equation}\label{cofx'}
   \co f(x')=\sum_i\lambda'_if(s_i)\;.
\end{equation}
Indeed, let $x'=\sum_j\mu_j t_j$ be any convex combination with
$t_j\in M$. Then  we can find a small number $\varepsilon>0$ such
that $(\lambda_i-\varepsilon\lambda'_i)\geq0$ for all $i$. Hence
\begin{displaymath}
   x=\sum_i(\lambda_i-\varepsilon\lambda'_i)s_i
       +\sum_j\varepsilon\mu_j t_j
\end{displaymath}
is a convex combination of elements from $M$ representing $x$. But
since the decomposition using only the $s_i$ is optimal, we have
\begin{displaymath}
   \sum_i(\lambda_i-\varepsilon\lambda'_i)f(s_i)
       +\sum_j\varepsilon\mu_j f(t_j)
    \geq\sum_i\lambda_if(s_i)\;
\end{displaymath}
From this we immediately get the claimed optimality of
$x'=\sum_j\lambda'_j s_j$.

These remarks are especially useful for the case of entanglement
of formation, for any mixed state the optimizing convex
decomposition necessarily involves several terms. Hence any
computation of an entanglement of formation immediately extends to
a larger class of states. Therefore, it is of great interest not
only to get the value of the entanglement of formation for a given
mixed state, but also to find the set of pure states solving the
variational problem defining $\EoF$

The symmetric situation studied in this paper is extreme in this
regard: The minimizing sets are always complete orbits of the
symmetry group. Therefore we get a fairly large set of
non-symmetric mixed states for which the computations below also
give the exact value of $\EoF$.

\subsection{Results for $G=UU$}
In this subsection we will apply the general method to computing
the entanglement of formation for the states of Example~1.

In the first step we have to determine the set $M_f$ of vectors
$\Phi\in\H\otimes\H$ such that $\braket{\Phi}{\flip\Phi}=f$. In
terms of the vector components $\Phi_{ij}$ we get
\begin{equation}\label{f-Phi}
  \braket{\psi}{\F \psi}
  =\sum_{ij}\Phi_{ij}\overline{\Phi}_{ji}\;.
\end{equation}
On the other hand, the reduced density operator has components
$\rho_{ij}=\sum_k\Phi_{ik}\overline{\Phi}_{jk}$ or, in matrix
notation, $\rho=\Phi\Phi^*$. Here we may introduce a
simplification due to $U\otimes U$ symmetry, by choosing $\rho$
diagonal. Note, however, that we can {\it not} choose the
restriction to the second system, i.e., $\Phi^T \bar \Phi$ to be
diagonal at the same time without loss of generality. In any case,
the eigenvalues of $\rho$ become
$\rho_{ii}=\sum_k\abs{\Phi_{ik}}^2$. Hence the pure state
entanglement of $\Phi$, which by (\ref{Epure}) is the entropy of
$\rho$ is
\begin{equation}\label{entUU}
  E(\ketbra\Phi\Phi)
     =\sum_i\eta\Bigl(\sum_k\abs{\Phi_{ik}}^2\Bigr)\;,
\end{equation}
where $\eta$ is the entropy function from (\ref{eta}).

For analyzing the variational problem it is useful to consider the
contributions of each pair of variables $\Phi_{ij}$ and
$\Phi_{ji}$, and of each diagonal element $\Phi_{ii}$ separately.
The weights of these contributions are
\begin{eqnarray}\label{pairweight}
  \lambda_{ij}&=&\abs{\Phi_{ij}}^2+\abs{\Phi_{ji}}^2\;,
               \quad\text{for}\ i<j\\
  \lambda_{ii}&=&\abs{\Phi_{ij}}^2\;.
\end{eqnarray}
The normalized contribution of one such pair or diagonal element
to $f$ is
\begin{eqnarray}\label{pairf}
  f_{ij}&=&\lambda_{ij}^{-1}\
            2\Re\bigl({\Phi_{ij}}\overline{\Phi_{ji}}\bigr)\;,
               \quad\text{for}\ i<j\\
  f_{ii}&=&1\;,\quad\text{so that }\\
      f &=& \sum_{i\leq j}\lambda_{ij}f_{ij}\;.
\end{eqnarray}
Similarly, we can write the probability distribution
$\rho_{11},\ldots,\rho_{dd}$ as a convex combination of
probability distributions with respective entropies
\begin{eqnarray}\label{paireta}
  s_{ij}&=&H_2\Bigl(\lambda_{ij}^{-1}\abs{\Phi_{ij}}^2\Bigr)\;,
               \quad\text{for}\ i<j\\
  s_{ii}&=&0\;,\label{singleeta}
\end{eqnarray}
where we have used the abbreviation $H_2(p)=\eta(p)+\eta(1-p)$ for
the entropy of a two point probability distribution $(p,1-p)$. By
concavity of the entropy we have
\begin{equation}\label{pairent}
    E(\ketbra\Phi\Phi)\geq\sum_{i\leq j}\lambda_{ij}s_{ij}\;.
\end{equation}
To find the lower bound on $s_{ij}$ given $f_{ij}$ is just another
instance of the variational problem we are solving, albeit with
the considerable simplification that only one off-diagonal pair of
components of $\Phi$ is non-zero. This leaves the following
problem:
\begin{quote} Given two complex variables $x,y$ with the constraint $\abs
x^2+\abs y^2=1$, with $2\Re(x\overline y)=f$, minimize $s=H_2(\abs
x^2)$.
\end{quote}
Since $s$ is monotonely increasing in $\abs x^2$ form $0$ to
$1/2$, this is equivalent to minimizing $\abs x^2$, given $f$. The
pairs $(\abs x^2,f)$ compatible with the constraints form the
convex set
\begin{displaymath}
\Set\big{(\lambda,f)}{\;\abs f\leq2\sqrt{\lambda(1-\lambda)};\
0\leq\lambda\leq1}\;.
\end{displaymath}
From this we get the minimal admissible $\abs
x^2=(1-\sqrt{1-f^2})/2$ in the above two variable variational
problem. Hence
\begin{equation}\label{eps2UU}
  s_{ij}\geq\varepsilon_2(f_{ij})
     = H_2\Bigl(\frac12\;\bigr(1-\sqrt{1-f_{ij}^2}\;\bigl)\Bigr)\;.
\end{equation}
This function $\varepsilon_2$ can be shown to be convex by
explicitly computing the second derivative and expanding
logarithms in a power series. Combining the bounds
(\ref{pairent}), (\ref{singleeta}), and (\ref{eps2UU}) with the
convexity of $\varepsilon_2$, we get
\begin{eqnarray*}
   E(\ketbra\Phi\Phi)
     &\geq&\sum_{i< j}\lambda_{ij}\;\varepsilon_2(f_{ij})\\
     &\geq&\varepsilon_2\Bigl(\sum_{i< j}\lambda_{ij}\;f_{ij}\Bigr)\\
     &=&\varepsilon_2\bigl(f-\sum_i\lambda_{ii}\bigr)\;.
\end{eqnarray*}
Now suppose that $f\geq0$. Then we can choose just a single
diagonal entry $\Phi_{ii}$ to be non-zero, and find
$E(\ketbra\Phi\Phi)=0$, which is clearly the minimum. However, if
$f<0$ the last equation shows that letting any diagonal entry
$\Phi_{ii}\neq0$ decreases the argument of $\varepsilon_2$ further
in a range where this function is monotonely decreasing. Hence the
optimum is choosing all $\Phi_{ii}=0$, and allowing only two
non-zero components $\Phi_{ij}$ and $\Phi_{ji}$ for some $i\neq
j$. This concludes the computation of $\EoF$ for $UU$-invariant
states (see summary below).

However, as noted in Section~\ref{sec:ext}, knowledge of the
minimizers for $\varepsilon$ automatically leads to an extension
of the computation to some non-invariant states. Let $x,y$ be a
solution of the two variable variational problem with
$f=\tr(\rho\flip)$. Then the minimizing vector is of the form
\begin{equation}\label{shuffle}
 x\ket{12}+y\ket{21}
   = \bigl(x\idty+y\flip\bigr)\ket{12}\;.
\end{equation}
All $U\otimes U$-translates of this vector will do just as well
and appear in the minimizing decomposition of the $UU$-invariant
state. Hence all convex combinations of the density operators
\begin{displaymath}
  \bigl(x\idty+y\flip\bigr)(U\otimes U) \ket{12}
  \bra{12}(U\otimes U)^*\bigl(x\idty+y\flip\bigr)^*
\end{displaymath}
with fixed $x,y$, and arbitrary $U$, have the same $\EoF$. For
determining these  convex combinations we can drop the outer
factors, and afterwards shift the operators found with
$(x\idty+y\flip\bigr)\in G'$. Let
\begin{equation}\label{FzeroFace}
  {\cal F}=\co\Set\big{(U\otimes U)\ket{12}\bra{12}
                    (U\otimes U)^*}{U\ \text{unitary}}\;.
\end{equation}
Clearly, every operator in ${\cal F}$ is a separable density
operator with flip expectation zero. Conversely, any operator
$\widetilde\rho$ with these properties may be decomposed into pure
product states $\ketbra{\phi\otimes\psi}{\phi\otimes\psi}$. These
must also have flip expectation zero, which means that
$\phi\perp\psi$, so that there is a unitary $U$ with
$\phi\otimes\psi=(U\otimes U)\ket{12}$. Consequently
$\widetilde\rho\in{\cal F}$.

Hence in order to determine whether for a given $\rho$ we can
compute $\EoF(\rho)$, we transform it to $\widetilde\rho$ by the
appropriate $(x\idty+y\flip)^{-1}$, and then test the separability
of $\widetilde\rho$.

\begin{figure}
\begin{center}
\epsfxsize=8.5cm \epsffile{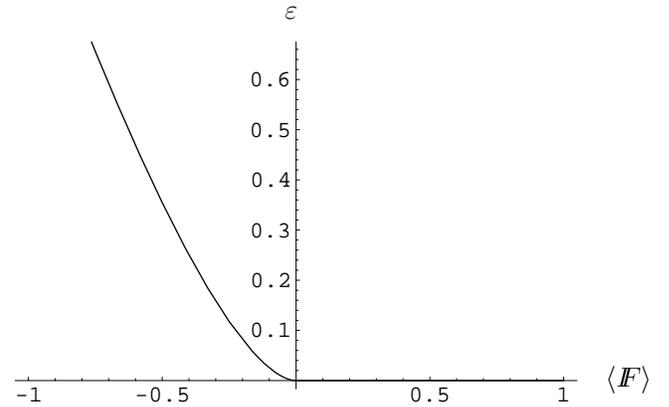}
\caption{ $\eps$-function for $\UU$ invariant states. }
\label{figUU}
\end{center}
\end{figure}

Let us summarize:
\begin{itemize}
\item
For the $U\otimes U$-invariant state $\rho$ with\newline
$\tr(\rho\flip)=f\leq0$, we have
\begin{equation}\label{eofUU}
  \EoF(\rho)=H_2\Bigl(\frac12\;\Bigr(1-\sqrt{1-f^2}\;\Bigl)\Bigr)\;,
\end{equation}
independently of the dimension $d$ of the underlying Hilbert
space. When $f\geq0$, the state $\rho$ is separable, and
$\EoF(\rho)=0$.
\item Let $\rho$ be a (not necessarily invariant) density operator
with $\tr(\rho\flip)=f$ and $-1<f<0$. Then with suitably chosen
$\alpha,\beta\in\Rl$

\begin{equation}\label{rhoUUtil}
  \widetilde\rho=(\alpha\idty+\beta\flip)^*\rho(\alpha\idty+\beta\flip)
\end{equation}
is a density operator with $\tr(\widetilde\rho\flip)=0$. Suppose
that $\widetilde\rho$ is separable. Then formula~(\ref{eofUU})
also holds for $\rho$.
\end{itemize}

\subsection{Results for $G=U\bar U$}
The computation of the entanglement of formation for Example~2 is
already known~\cite{TerVol}. The minimizing pure states are of the
form \beq
 \bigl(x\idty+y\flipt\bigr)\ket{11}\;,
\eeq
with real $x,y$.

The extension to non-invariant states works
in principle similar to the $\UU$-case, but for $d>2$ it is
getting a bit more complicated, because the $\eps$-function is
not convex anymore.

\begin{figure}
\begin{center}
\epsfxsize=8.5cm \epsffile{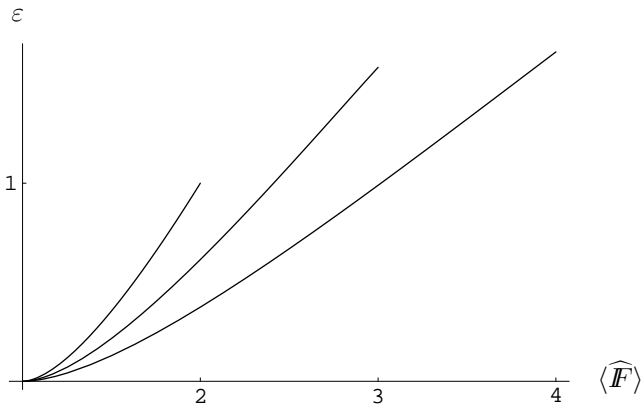} \caption{ $\eps$-function for
$\UUb$ invariant states for $d=2,3,4$. The functions are not convex near the
right endpoint for $d\geq 3$.} \label{figUUb}
\end{center}
\end{figure}
\begin{itemize}
\item
For the $U\otimes \Ubar $-invariant state $\rho$ with\newline
$\tr(\rho\flipt)=f \geq \frac{1}{d}$, we have
\begin{equation}\label{eofUbU}
  \EoF(\rho)=\co \Bigl(H_2(\gamma)+(1-\gamma) \log(d-1) \Bigr)\;,
\end{equation}
with $\gamma=\frac{1}{d^2} \left( \sqrt{f}+\sqrt{(d-1)(d-f)}
\right)^2$. For $d
> 2$ we need also to compute the convex hull. When
$f<\frac{1}{d}$,
 the state $\rho$ is separable, and $\EoF(\rho)=0$.
\item Let $\rho$ be a (not necessarily invariant) density operator
with $\tr(\rho\flipt)=f$, $1<f<d$ and $\co(\eps(f))=\eps(f)$.
Then with suitably chosen
$\alpha,\beta\in\Rl$
\begin{equation}\label{rhoUbUtil}
  \widetilde\rho=(\alpha\idty+\beta\flipt)^*\rho(\alpha\idty+\beta\flipt)
\end{equation}
is a density operator with $\tr(\widetilde\rho\flipt)=1$. Suppose
that $\widetilde\rho$ is separable. Then formula (\ref{eofUbU})
also holds for $\rho$.

\item If $f$ satisfies $\co(\eps(f))< \eps(f)$ , the convex hull has a
flat section between $f_1<f<f_2$ where $f_1, f_2$ are the two end
points of the flat piece satisfying  $\co(\eps(f_{1/2}))=
\eps(f_{1/2})$. We can always find a convex decomposition of
$\rho$ in two states with expectation values $f_1, f_2$. If now
the above procedure works for these two states, then we have found
an optimal decomposition for $\rho$ and can easily compute the
entanglement of formation.

\end{itemize}

\subsection{Results for $\OO$-invariant states}

\begin{figure}
 \begin{center} \epsfxsize=6.5cm \epsffile{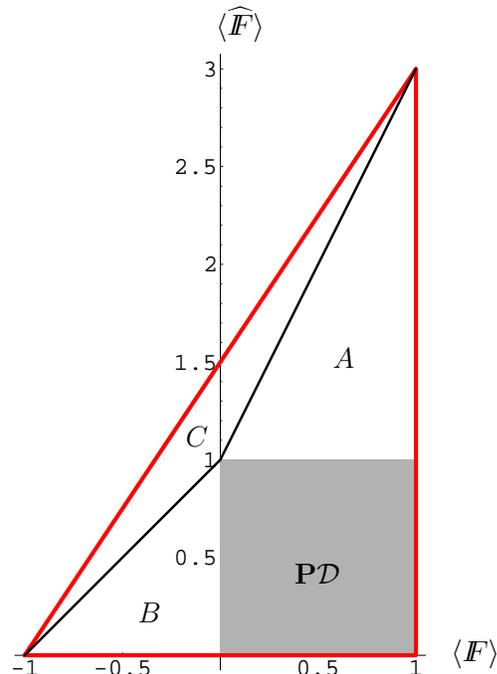}
\caption{The state space for
$OO$-invariant states seems to split naturally in four regions.
The separable square $\pr\sep$ and the three triangles $A,B,C$.}
\label{OO}
 \end{center}
 \end{figure}

Here the extension method of Section~IV.B turns out to do much of
the work.  The state space, plotted in Figure~\ref{OO}, is
separated in four regions. The separable square and the three
triangles $A, B, C$.
 In order to apply the extension method to the $UU$-invariant states,
we have to see which states can be written as
$\rho=(x\idty+y\flip)\til\rho(x\idty+y\flip)^*$, with $\til\rho$ a
separable state with flip expectation zero. If we take for
$\til\rho$ any state at the left edge of the separable square, it
is clear that we will get an $\OO$-invariant state again. The
explicit computation shows that with this method we get
$\EoF(\rho)$ in the full triangle $B$. Note that by this
construction $\EoF(\rho)$ depends only on the expectation
$\langle\flip\rangle$, and not on $\langle\flipt\rangle$ or the
dimension $d$.  Employing similarly the extension method for
$\UUb$ we find $\EoF(\rho)$ in the triangle $A$, getting a
function depending only on $\langle\flipt\rangle$ and the
dimension, but not on $\langle\flip\rangle$.

\subsection{Results for Bell-States} The Bell-States were one of
the first classes for which entanglement of formation could be
calculated \cite{EoF}. Of course, our method reproduces this
result, albeit with a more economical decomposition. This is a
feature shared with the Wootters formula\cite{Wootters}. It is a
natural question whether the extension method, applied in this
basic example, reproduces the Wootters formula. However, it turns
out that one gets the result only on state manifolds of lower
dimension. We also did not succeed in finding another group of
local symmetries, which would give Wootter's formula in full
generality.

\section{Relative Entropy of Entanglement}
\subsection{Simplified computation}

Symmetry simplifies the computation of the relative entropy of
entanglement dramatically: it reduces the variation in (\ref{ere})
from a variation over all separable states $\sigma\in\sep$ to
those, which are also $G$-invariant. i.e., when $\rho=\pr\rho$, we
have
\begin{equation}\label{eres}
  \ERE(\rho)
     =\inf\Set\big{\SS(\rho,\sigma)}{\sigma\in \pr\sep}\;.
\end{equation}
The only ingredients of the proof are the convexity of
$\sigma\mapsto\SS(\rho,\sigma)$, the invariance of relative
entropy under (local) unitary transformations of both its
arguments, and that $\sep$ is a convex set invariant under local
unitaries. Indeed, the properties of $\sep$ imply that for any
$\sigma$ in the full variational problem,
$\pr\sigma\in\pr\sep\subset\D$ is also a legitimate argument, and
the convexity properties of $\SS$ show that this cannot increase
$\SS(\rho,\sigma)$. Hence the variation may be restricted as in
(\ref{eres}). We have listed the ingredients of the proof so
explicitly, because many variations of $\ERE$ may be of
interest. For example, the ``distance'' function relative entropy
can be replaced by a host of other functions, like norm
differences of any kind. The set $\sep$, too, may be replaced, for
example by the set of ppt-states, as suggested by Rains
\cite{Rains}, who also made similar use of symmetry.

A second simplification concerns the computation of
$\SS(\rho,\sigma)$ itself, when both arguments are $G$-invariant.
We have seen that $G$-invariant states can be considered as states
on the commutant algebra $G'$. Now the relative entropy is defined
for pairs of states on arbitrary C*-algebras \cite{Petz}, and the
form~(\ref{relent}) involving density matrices is only the special
form valid for a full matrix algebra. Since $\pr$ is a conditional
expectation onto $G'$, the result does not depend \cite{Petz} on
whether we compute the relative entropy via density matrices, or
for the corresponding abstract linear functionals on $G'$. Without
going into the details for general algebras $G'$ here, let us see
how this helps in the case when $G'$ is abelian, as in most of our
examples.

Suppose $p_\alpha$, $\alpha=1,\ldots,N$ are the minimal
projections of $G'$, and denote by $\omega_\alpha=(\tr
p_\alpha)^{-1}p_\alpha$ the extremal density matrices of $\pr\st$.
Then every $\rho\in\pr\st$ has a unique representation as a convex
combination
\begin{equation}\label{rhospec}
  \rho=\sum_\alpha\rho_\alpha\omega_\alpha
      =\sum_\alpha\frac{\rho_\alpha}{\tr p_\alpha}\ p_\alpha\;,
\end{equation}
where the second expression is at the same time the spectral
resolution of $\rho$. If we compute the von Neumann entropy
$-\tr(\rho\log\rho)$ from this, we find a dependence of the result
not only on the expectations $\rho_\alpha=\tr(\rho p_\alpha)$, but
also on the multiplicities $\tr(p_\alpha)$, as is quite familiar
from statistical mechanics. On the other hand, the fact that
relative entropy can be defined for states on abstract algebras
shows that no such dependence can occur for relative entropies.
Indeed, the terms involving $\log\tr(p_\alpha)$ from $\rho$ and
$\sigma$ cancel, and we get
\begin{equation}\label{relentab}
  \SS(\rho,\sigma)=\sum_\alpha \rho_\alpha
      \Bigl(\log(\rho_\alpha)-\log(\sigma_\alpha)\Bigr)\;,
\end{equation}
where $\rho_\alpha$ and $\sigma_\alpha$ are the respective
expectations of $p_\alpha$.

A typical application is the observation that for $UU$-invariant
states the expression for the relative entropy of entanglement can
be written down in terms of the $\tr(\rho F)$, independently of
the dimension $d$ of the underlying Hilbert spaces.

For $\UU$ and $\UUb$-invariant states the sets  of separable
states are just intervals, and the definition of relative entropy
of entanglement requires a minimization over this interval.
However, due to a general property of the relative entropy, the
convexity in both arguments, it is clear that the minimum is, in
fact always obtained at the endpoint: if $\rho$ is the state whose
entanglement we want to calculate, and $\sigma$ is the minimizing
separable state, convexity implies
\begin{eqnarray*}
  \SS(\rho,\lambda \sigma+(1-\lambda)\rho)
    &\leq& \lambda\SS(\rho,\sigma)+(1-\lambda)\SS(\rho,\rho)\\
    &=&\lambda\SS(\rho,\sigma)\;.
\end{eqnarray*}
Hence if there were any separable state on the straight line
segment connecting $\rho$ and $\sigma$, it would give a strictly
lower minimum, contradicting the minimality of $\sigma$.

\begin{figure}
\begin{center}
\epsfxsize=8.5cm \epsffile{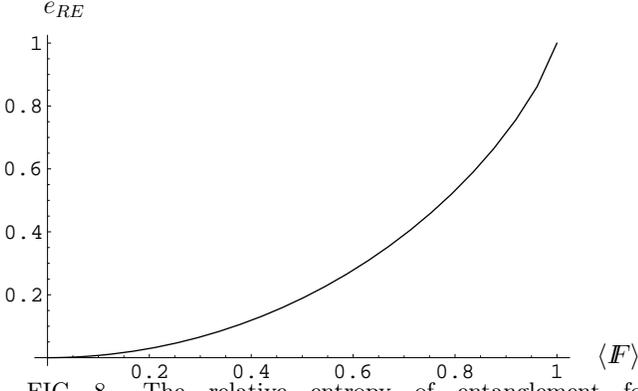} \caption{The relative entropy of
entanglement for UU-invariant states. } \label{RE}
\end{center}
\end{figure}

For $UU$ the boundary separable state has $\tr(\sigma\flip)=0$,
i.e., gives equal weight to the minimal projections. We have to
compute the relative entropy with respect to a state with
probabilities $(1\pm f)/2$, i.e., the function
\begin{equation}\label{eREf}
  e_{\rm RE}(f)=
      \log2-S  \left(\frac{1+f}2,\frac{1-f}2 \right)\; ,  
\end{equation}
where we denote by $S(p_1,\ldots,p_n)=-\sum_kp_k\log p_k$ the
entropy of a probability vector $(p_1,\ldots,p_n)$. This function
is plotted in Figure~\ref{RE}, and determines the relative entropy of
entanglement of $\UU$-symmetric states $\rho$ via
\begin{equation}\label{EREUU}
  \ERE(\rho)=e_{\rm RE}\bigl(\tr(\rho\flip)\bigr)\;.
\end{equation}

Similarly, the boundary point $\sigma$ of $\D$ for $\UUb$
invariant states is given by $\tr(\sigma\flipt)=1$.  For general
$\ft=\tr(\rho\flipt)$ the minimal projections have weights $\ft/d$
and $(1-\ft/d)$. Hence for $\UUb$-symmetric $\rho$, we have
$\ERE(\rho)=0$ for $\ft\leq1$, and
\begin{equation}\label{eREft}
   \log d-(1-\frac\ft d)\log(d-1)-S\Bigl(\frac\ft d,1-\frac\ft d\Bigr)
\end{equation}
otherwise. For comparison with the results of \cite{TerVol}, note
that $\ft/d$ is the so called {\it maximally entangled fraction}
of $\rho$.

Now we look at $OO$-invariant states. The state space and the
separable states are drawn in Figure~\ref{OO}. First we look at
the state with the coordinates $(1,3)$, which is a maximal
entangled state. The separable states, that are minimizing the
relative entropy for this state, are the states on the whole line
connecting the points $(0,1)$ and $(1,1)$. But now we can find the
minimizing separable for any state in the whole triangle $A$. We
just have to draw the straight line connecting the coordinates of
a given states with the point $(1,3)$. The intersection with the
border of $\pr \D$ is then a minimizer for $(1,3)$ and by the
properties of the relative entropy of entanglement also the
minimizer for all states on the connecting line. The same
argumentation works for the edge point $(-1,0)$ and the separable
border between $(0,0)$ and $(0,1)$ giving us all minimizers for
the triangle $B$. The whole triangle $C$ has the same minimizer,
namely $(0,1)$.

\subsection{Counterexample to additivity}
To find a counterexample to the additivity of the relative entropy
of entanglement we use the group introduced in Example~7 . We also
know that additivity will hold for any states where one of the two
independently prepared states is separable. So in our example we
can restrict to the area, where both expectation values of $F_1$ and
$F_2$ are negative.

\begin{figure}
\begin{center}
\epsfxsize=8.5cm \epsffile{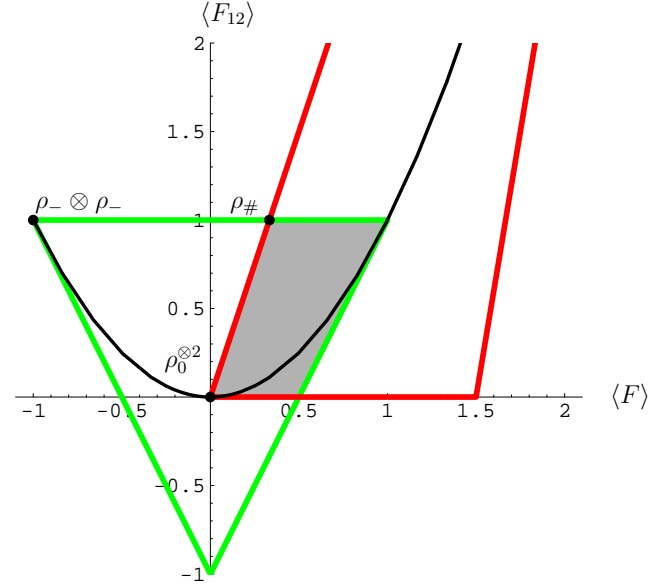} \caption{ State space for
$UUVV \& F$-invariant states for $d=3$. } \label{figUVUV}
\end{center}
\end{figure}

 For simplicity we increase the group with $\F_{\rm
Alice} \otimes \F_{\rm Bob}$ leading us to a smaller commutant
only spanned by $\idty \otimes \idty, \F \otimes \F, \idty \otimes
\F + \F \otimes \idty$. As coordinate system we use the
expectation values of \bea F&=&\frac{1}{2}( \idty \otimes \F + \F
\otimes \idty) \\ F_{12}&=&\F \otimes \F . \eea The state space is
drawn in Figure~\ref{figUVUV}.
It is just the intersection of the state space of the original
group (see Figure~\ref{fig4}) with the plane given by $\langle F_1 \rangle=\langle F_2 \rangle =\langle F \rangle$.
The product states, in the sense of additivity, are given by the
line $(\langle F \rangle ,\langle F \rangle ^2)$.

The counterexample we want to look at, is the state referring to
the coordinates $(-1,1)$, which is given by $\rho_- \otimes
\rho_-$ where $\rho_-$ denotes the normalized projection on the
antisymmetric subspace of $\Cx^d \otimes \Cx^d$. From
Equation~(\ref{EREUU}) we know the relative entropy of
entanglement for $\rho_-$ to be $\log 2$ independent from the
dimension $d$. The minimizing state was  the state with flip
expectation value equal to zero now denoted as $\rho_0$. So the
expected minimizer for the tensor product would be  $\rho_0
\otimes \rho_0$ located on the quadratic product line with the
expectation values $(0,0)$. This one  gives us the expected value
of $\log 4$ for the relative entropy. Now we calculate the
relative entropy between $\rho_{-}^{\otimes 2}=\rho_- \otimes
\rho_-$ and $\rho_\#$.
\bea \SS(\rho_-^{\otimes 2},\rho_\#)&=&
\tr(\rho_-^{\otimes 2} \log \rho_-^{\otimes 2} -\rho_-^{\otimes 2}
\log \rho_\#)\\ &=&\tr(\rho_-^{\otimes 2} \log \rho_-^{\otimes 2}
-\rho_-^{\otimes 2} \log \frac{d-1}{2d}\rho_-^{\otimes 2})\\
&=&-\log \frac{d-1}{2d}=\log 4 -\log \frac{2(d-1)}{d} \;. \eea
Indeed
the minimum must be attained on the line connecting $\rho_0
\otimes \rho_0$ and $\rho_\#$ and it can easily be verified, that
the minimum always is attained on $\rho_\#$. For $d=2$ the whole
line gives the same value and although there exits minimizer not
belonging to the product space, additivity holds. For $d>2$ the
expectation values of state $\rho_\#$ given by $(\frac{1}{d},1)$
shift near to the $F_{12}$ axis and from a geometrical point of
view closer to $\rho_- \otimes \rho_-$. Although the relative
entropy  is not a real kind of geometrical measure this
intuition did not fail. In these cases the additivity is violated
with an amount of $\log \frac{2(d-1)}{d}$. For very high dimension
$d$ we get the really surprising result $E_{RE}(\rho_-\otimes
\rho_-) \simeq E_{RE}(\rho_-)$.

\section{Concluding Remarks} We have concentrated on just two
basic entanglement measures. Clearly, there are many more, and for
many of them the computation can be simplified for symmetric
states. Among these measures of entanglement are the ``best
separable approximation'' of a state \cite{AnLe}, the trace norm
of the partial transpose \cite{HHH}, the base norm associated with
$\sep$ (called cross norm in \cite{Rudolph} and absolute
robustness in \cite{robust}). For distillible entanglement we
refer to the recent paper of Rains \cite{Rains2}. Similarly, there
is a lot of work left to be done carrying out the programme
outlined in this paper for all the groups of local symmetries
listed  in Section \ref{hier}.

\section*{Acknowledgement}
Funding by the European Union project EQUIP (contract
IST-1999-11053) and financial support from the DFG (Bonn) is
gratefully acknowledged.


\begin{references}

\bibitem{Wer} R. F. Werner, Phys. Rev. A {\bf 40}, 4277 (1989).
\bibitem{Pop} S. Popescu, Phys. Rev. Lett. {\bf 72}, 797 (1994).
\bibitem{isotropic} M. Horodecki and P. Horodecki, Rhys. Rev. A {\bf
59}, 4206 (1999).


\bibitem{tripartite} T. Eggeling and R. F. Werner,
quant-ph/0003008.
\bibitem{tripartiteC} W. Dür, J.I. Cirac and R. Tarrach, Phys.
Rev. Lett. {\bf 83}, 3562 (1999).

\bibitem{Rains2} E. M. Rains, quant-ph/0008047.

\bibitem{ERE}  V. Vedral, M.B. Plenio, M.A. Rippin and P. L.
Knight, Phys. Rev. Lett. {\bf 78}, 2275 (1997).
\bibitem{EREW}  V. Vedral and M.B. Plenio, Phys. Rev.  A {\bf 57},
1619(1998).

\bibitem{EoF} C. H. Bennett, D. P. DiVincenzo, J. A. Smolin and W. K.
Wootters, Phys. Rev. A {\bf 54}, 3824 (1996).

\bibitem{Wootters} W. K. Wootters, Phys. Rev. Lett. {\bf 80}, 2245
(1998).
\bibitem{Uhlmann}In the present context this has also been called
the roof property of $f$.
 A. Uhlmann, Open Sys.\& Inf. Dyn. {\bf 5}, 209 (1998).
\bibitem{TerVol}B. Terhal and K.G.H. Vollbrecht,  Phys. Rev. Lett. {\bf
85}, 2625 (2000).
\bibitem{Rains} E. M. Rains, Phys. Rev. A {\bf 60}, 179 (1999).
\bibitem{Weyl} H. Weyl, {\it The Classical Groups}, (Princeton University,
1946).
\bibitem{Peres} A. Peres, Phys. Rev. Lett. {\bf 77}, 1413 (1996).
\bibitem{WV}K.G.H. Vollbrecht and R.F. Werner, J. Math Phys. {\bf
41}, 6772 (2000).
\bibitem{bicommutant}M. Takesaki, {\it Theory of Operator Algebras
I}, (Springer-Verlag 1979).
\bibitem{BeNar}F. Benatti and H. Narnhofer, quant-ph/0005126.

\bibitem{Petz} M. Ohya and D. Petz, {\it Quantum Entropy and Its
Use}, (Springer-Verlag 1993).

\bibitem{AnLe} M. Lewenstein  and A. Sanpera, Phys. Rev. Lett. {\bf 80}, 2261
(1998).

\bibitem{HHH}  M.Horodecki, P. Horodecki and. Horodecki,
Phys. Rev. Lett. {\bf 84}, 4260 (2000).



\bibitem{Rudolph} O. Rudolph,  J. Phys. A: Math. Gen. {\bf 33}, 3951
(2000).


\bibitem{robust} G. Vidal and R. Tarrach, Phys. Rev. A {\bf 59}, 141-155 (1999).




\end{references}
\end{document}